\documentclass[aps,prl,showpacs,floatfix,superscriptaddress,twocolumn]{revtex4-1}
\usepackage{amsmath,amssymb,physics}
\usepackage{verbatim,graphicx}

\usepackage{color}
\usepackage[all]{xy}
\usepackage{color}
\usepackage[all]{xy}

\newcommand{\be}{\begin{eqnarray}}
\newcommand{\ee}{\end{eqnarray}}

\newcommand{\bmat}{\left ( \begin{array}{cc} }
	\newcommand{\emat}{\end{array} \right ) }

\newcommand{\beq}{\begin{equation}}
\newcommand{\beqs}{\begin{equation*}}
\newcommand{\eeq}{\end{equation}}
\newcommand{\eeqs}{\end{equation*}}

\allowdisplaybreaks

\begin{document}

\title{Many-Body Localization in a finite-range Sachdev-Ye-Kitaev model}
\author{Antonio M. Garc\'\i a-Garc\'\i a}
\affiliation{Shanghai Center for Complex Physics, 
	Department of Physics and Astronomy, Shanghai Jiao Tong
	University, Shanghai 200240, China}
\email{amgg@sjtu.edu.cn}
\author{Masaki Tezuka}
\affiliation{Department of Physics, Kyoto University, Kyoto 606-8502, Japan}

\begin{abstract} 
We study the level statistics of a generalized Sachdev-Ye-Kitaev (SYK) model 
with two-body and one-body random interactions of finite range by exact diagonalization. Tuning the range of the one-body term, while keeping the two-body interaction sufficiently long-ranged, does not alter substantially the  spectral correlations, which are still given by the random matrix prediction typical of a quantum chaotic system. However a transition to an insulating state, characterized by Poisson statistics, is observed by reducing the range of the two-body interaction. Close to the many-body metal-insulator transition, we show that spectral correlations share all features previously found in systems at the Anderson transition and in the proximity of the many-body localization transition. Our results suggest the potential relevance of SYK models in the context of many-body localization and also offer a starting point for the exploration of a gravity-dual of this phenomenon.       
\end{abstract}

\maketitle
The claim \cite{basko2006,basko2006a}, based on analytical arguments but with uncontrolled approximations, that Anderson localization is stable to the presence of weak interactions has given a new impetus to the problem of the interplay of disorder and interactions \cite{fleishman1980,pikovsky2008,shepelyansky1993,shepelyansky1994}. A direct precursor of this result was the prediction \cite{altshuler1997,shepelyansky1993} of a metal-insulator transition in the Fock space at finite coupling in disordered strongly interacting quantum dots.

More recent research \cite{luitz2015,luitz2016,agarwal2015,bardarson2012,gopalakrishnan2015} has focused on the description of this new state of matter, where both interaction and disorder are important, loosely referred to as many-body localization \cite{basko2006,basko2006a,oganesyan2007}. Typical features of the insulating region, for short range interactions, include slow logarithmic growth of the entanglement entropy \cite{bardarson2012}, level statistics given by Poisson statistics \cite{oganesyan2007,aizenman2007}, zero dc conductivity  \cite{basko2006} and vanishing of the ac conductivity as a power-law \cite{gopalakrishnan2015}. By contrast, the metallic region close to the transition shares features with a Griffith phase \cite{agarwal2015} where diffusion is very slow and the growth of the entanglement entropy is power-law \cite{luitz2016}. Consistent with this anomalous sub-diffusion, level statistics, at least for the small volume accessible to numerical calculations, are close to that found in systems at the metal-insulator transition \cite{corentin2016}. With important exceptions \cite{wang2008,imbrie2016,imbrie2016a,aizenman2007,aizenman2009,aizenman2011}, most of these results are based on numerical calculations in relatively small lattices which make it difficult to discern which of these features will survive in the thermodynamic limit. 

For that reason, it would be interesting to have a simple toy model, with both interactions and disorder, which could reproduce most of the desired features of many body localization but still be more amenable to an analytical treatment. In principle this seems a hopeless task, however in the context of high energy physics a simple model for holography has been recently proposed, usually termed Sachdev-Ye-Kitaev model \cite{kitaev2015,jensen2016,maldacena2016,sachdev2015,sachdev1993,sachdev2010} which is solvable, at least in some region of parameters, despite being disordered and strongly coupled. In its simplest form, proposed by Kitaev \cite{kitaev2015}, it consists of $N$ Majorana fermions in zero spatial dimensions with infinite-range interactions though similar models have been used in nuclear physics, quantum chaos and condensed matter \cite{bohigas1971,bohigas1971a,french1970,french1971,mon1975,benet2003,kota2014,sachdev1993,parcollet1998,georges2001} for a long time. In the strong coupling limit, the SYK model is characterized by a finite entropy at zero temperature, the specific heat linear in temperature, the saturation of a recently proposed \cite{maldacena2015} bound on chaos and a density of low energy excitation which grows exponentially with energy \cite{maldacena2016,garcia2016,cotler2016,garcia2017}.
Moreover level statistics are well described by random matrix theory with deviations consistent with the existence of a Thouless energy in the system \cite{garcia2016,cotler2016,garcia2017,garcia2018} and 
the conductivity in higher dimensional generalizations of the model \cite{gu2016,jian2017,cenke2017,altman2017,davison2017} is finite. These are features expected of a field theory with a gravity dual but also of a strongly coupled disordered metal.

A natural question to ask is whether it is possible to deform the SYK model so that it undergoes a metal-insulator transition in Fock space. That would make possible not only to study the rich phenomenology of many-body localization in a simpler model but also to explore the existence of a gravity dual. As was mentioned previously \cite{altshuler1997,shepelyansky1993}, interacting quantum dots can undergo a metal insulator transition in sparse lattices like a Cayley tree. Some recent papers \cite{jian2017,cenke2017,song2017,chen2017,garcia2017} in the holography literature have already explored the stability of the metallic phase of the SYK model however the employed models have interactions of infinite range, so the transition is more of the chaotic-integrable type and therefore not related to Anderson localization induced by quantum coherence effects, or are defined in higher spatial dimensions in which case the simplicity of the model is lost.

 Here we address this question by a numerical level statistics analysis of a generalized SYK model with one-body and two-body finite range interactions. 
By tuning the range of the two-body interaction we have identified a metal-insulator transition in the spectrum.
As was expected, spectral correlations are well described by random matrix theory and Poisson statistics in the metallic and insulating regions respectively. Around the transition, we have found that level statistics are strikingly similar to that of systems at the Anderson transition \cite{shapiro1993,altshuler1988,schreiber1991,garcia2005,bertrand2016} and in interacting disordered metals close to the many-body localization transition. 
This is a strong indication that this generalized SYK model could be employed as a toy model in studies of many-body localization \cite{altshuler1997,basko2006,oganesyan2007,luitz2015,bertrand2016}. Next we introduce the model and its main features. 
\section{The model}
We study the following Hamiltonian $0+1$ space-time dimensions, 
 \begin{equation}\label{hami}
H \, =\sum_{1=i<j<k<l}^N {\tilde J}_{ijkl}(D) \, \chi_i \, \chi_j \, \chi_k \, \chi_l \, +\, {i } {\kappa}\sum_{1=i<j}^N {\tilde K}_{ij}(d) \, \chi_i \, \chi_j \, 
\end{equation}
where $\kappa > 0$, $\chi_i$ are $N$ Majorana fermions defined by
$
\{ \chi_i, \chi_j \} = \delta_{ij}$, ${\tilde J}_{ijkl}(D)= 0$ if ${\rm max}(i,j,k,l) - {\rm min}(i,j,k,l) = l - i \geq D$ and ${\tilde J}_{ijkl}(D) = J_{ijkl}$ otherwise. We can define a fractional distance by gradually removing the active bonds between two consecutive integer values of $D$. An analogous definition applies to ${\tilde K}_{ij}(d)$: ${\tilde K}_{ij} = 0$ if $|i - j| \geq d$ and ${\tilde K}_{ij} = K_{ij}$ for $|i -j| < d$, so that for $d = 2$, only $K_{i,i+1}\vert_{i=1}^{N-1}$ are nonzero.
The couplings $J_{ijkl}$ and $K_{ij}$ are real random variables with distributions
\begin{equation}
P(J_{ijkl}) \, = \, \sqrt{\frac{N^3}{12 \pi}} \exp\left( - \, \frac{N^3J_{ijkl}^2}{12} \right) \, ,
\end{equation}  
\begin{equation}
P(K_{ij}) \, = \, \sqrt{\frac{N}{2 \pi}} \exp\left( - \, \frac{NK_{ij}^2}{2} \right) \, .
\end{equation} 

We note that Eq.(\ref{hami}), as was mentioned previously, is qualitatively similar to a toy model for the metal-insulator transition in Fock space at finite $N$ in quantum dots \cite{altshuler1997}. The main differences, putting aside that Ref.\cite{altshuler1997} studies Dirac rather than Majorana fermions, is that the interaction in \cite{altshuler1997} is not of this type but rather is constrained to a Cayley tree and the one-body term is restricted to nearest neighbors.  

As was mentioned previously, the SYK model with infinite range interactions is strongly chaotic. It was recently found \cite{garcia2017a} that chaoticity persists in the bulk of the spectrum for the values of $\kappa$ investigated in the paper, in the limit of infinite range interactions, so any transition induced by reducing the range of the interaction will likely be of the metal-insulator type. 
We study spectral correlations which are known to be a powerful probe  of the presence and characterization of a broad type of disordered and chaotic systems, from weakly disordered metals to Anderson insulators and critical chaotic systems \cite{altshuler1988,shapiro1993,braun1995,mirlin2000,garcia2005}.   
For comparison to the universal results of random matrix theory in the metallic region we need to determine the global symmetries of our model. In the SYK model without the one-body term, it is known that  depending on $N$ \cite{you2016,garcia2016} the Clifford algebra can admit real, complex or quaternionic representations that label different universality classes. However the one-body term in Eq.~(\ref{hami}) breaks the time reversal invariance so we expect that global symmetries belong in all cases to the broken time reversal invariance universality class described by the Gaussian Unitary Ensemble (GUE) \cite{mehta2004}.

\section{Level Statistics}
We investigate spectral correlations in the spectrum of Eq.~(\ref{hami}) by using exact diagonalization techniques. For a given set of parameters $\kappa, N$ we have obtained at least $10^6$ eigenvalues.
We note that the mean level spacing is, in general, energy dependent so for a meaningful comparison between different spectral intervals of the same system, the unfolding of the spectrum is required so that the mean level spacing is the same, unity for convenience, across the spectrum. 
 
We first compute the level spacing distribution $P(s)$ that probes the system dynamics for times of the order of the Heisenberg time, the inverse of the mean level spacing $\Delta$ in units of $\hbar$. It is defined as the probability $P(s)$ to find two consecutive eigenvalues $E_{i}, E_{i+1}$ at a distance $s = (E_{i+1}-E_{i})/\Delta$. 

For an insulator, we expect the Poisson statistics
$
P_\mathrm{P}(s) = e^{-s}$
while for a disordered metal or a quantum chaotic system $P(s)$ is given by the random matrix result for the GUE, which is well approximated by the so called Wigner--Dyson statistics \cite{mehta2004,DietzHaake1990}, 
\begin{equation}
P_\mathrm{GUE}(s) \approx P_\mathrm{W}(s) = \frac{32}{\pi^2}s^2{\rm e}^{-4 s^2/\pi}.
\label{eq:wd}
\end{equation}

\begin{figure}
	\centering
	\resizebox{0.5\textwidth}{!}{\includegraphics{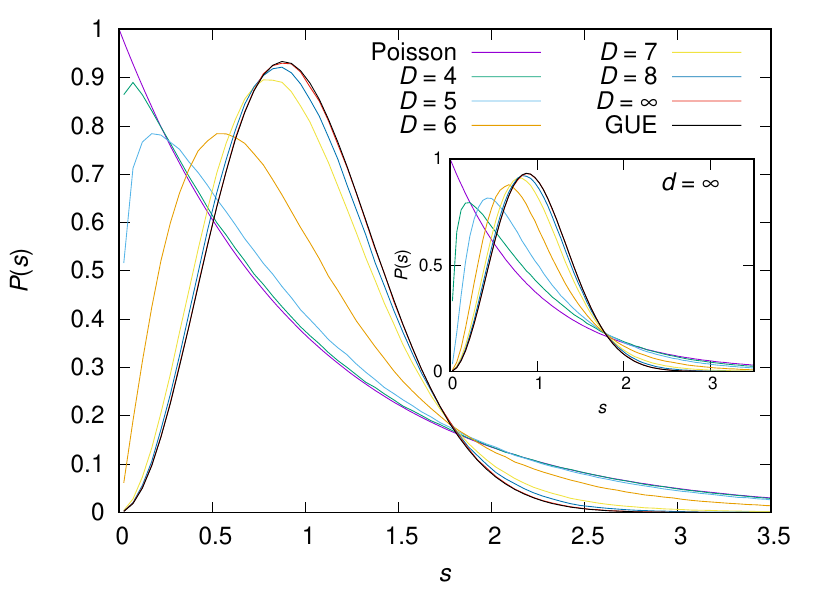}}
	\resizebox{0.5\textwidth}{!}{\includegraphics{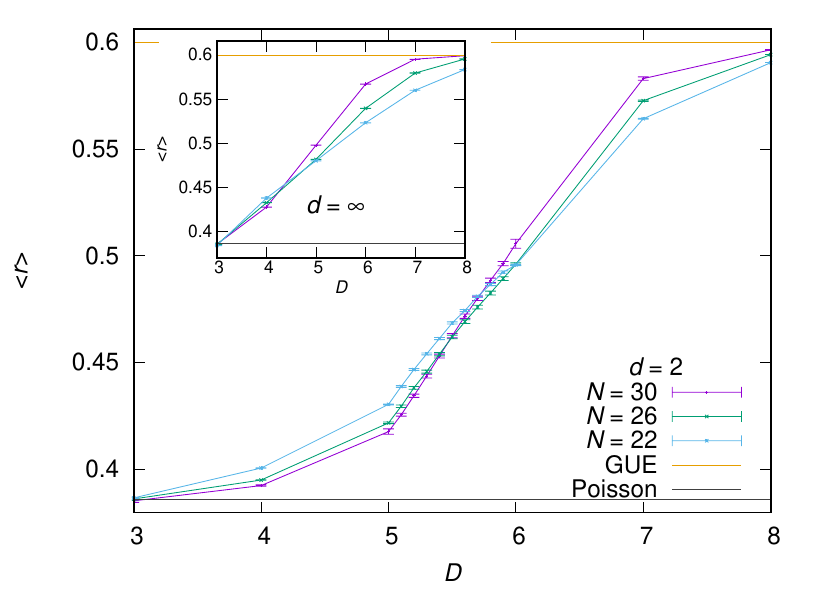}}
	\caption{
		Upper: Level spacing distribution $P(s)$ for $N = 30$, $\kappa = 1$, $d = 2$ and different cutoffs $D$.  A tenth order polynomial fitting has been employed to unfold the entire spectrum.  We observe a transition from the GUE prediction \cite{DietzHaake1990} to Poisson statistics $P_\mathrm{P}(s) = e^{-s}$ as $D$ decreases at $D_\mathrm{c} \approx 5.6$.
 Inset: Qualitatively similar results are observed for $d \to \infty$. Therefore reducing the range of the one-body random term does not affect qualitatively level statistics. Lower: Average adjacent gap ratio  $\langle r \rangle$ as a function of $D$ for different $N$'s and $d=2$. \cite{notefractionalD} The observed crossing point is a signature of a metal-insulator transition. Inset: Similar results are observed for $d = \infty$. }
	\label{fig3}%
\end{figure}
 Results for 
 $P(s)$, depicted in the upper plot of Fig.~\ref{fig3}, clearly indicate that, for sufficiently large $D$, Eq.~(\ref{eq:wd}) is in excellent agreement with the SYK model for $\kappa = 1$, even if the one-body interaction is restricted to nearest neighbors (main figure). Indeed results for an infinite range one-body interaction (inset) are similar, which suggests that level statistics are not very sensitive to the one-body interaction in this range of parameters.
However for $D$ sufficiently small we observe a transition to what it looks an insulating state characterized by Poisson statistics.

In order to clarify whether a true transition takes place we carry out a finite size scaling analysis \cite{shapiro1993}. For that purpose, we employ the adjacent gap ratio \cite{luitz2015,oganesyan2007}, 
\begin{equation}
r_i = \frac{\min(\delta_i, \delta_{i+1})}{\max(\delta_i, \delta_{i+1})} 
\label{eq:agr}
\end{equation}
for an ordered spectrum $E_{i-1} < E_i < E_{i+1}$ as the scaling variable where $\delta_i = E_i - E_{i-1}$.
The average adjacent gap ratio for a Poisson distribution is $\left\langle r \right\rangle_\mathrm{P} = 2\ln(2) - 1 \approx 0.386$. For the GUE it is  $\left\langle r \right\rangle_\mathrm{GUE}\approx 0.5996$ \cite{atas2013}. This quantity has the advantage over other scaling variables, like the moments of $P(s)$, that no unfolding of the spectrum is necessary. This reduces the probability of systematic errors. The crossing point, where the adjacent gap ratio becomes almost $N$ independent, occurs at $D = D_\mathrm{c} \approx 5.6$ in the lower plot of Fig. \ref{fig3}, which is a clear indication of the existence of a metal-insulator transition. We note that due to the relatively small range of $N$'s that we can investigate numerically, the value of $D_\mathrm{c}$ will, in view of our results, likely shift to a higher value in the large $N$ limit.

Having established the existence of the metal-insulator transition, we now characterize it by the study of both short-range and long-range spectral correlations at $D \sim D_\mathrm{c}$. We aim to clarify whether it shares features with the Anderson metal-insulator transition \cite{shapiro1993} for disordered non-interacting systems or, for interacting disordered metals, close to the many-body localization transition \cite{bertrand2016}. In both cases it is well known that the system becomes scale invariant and well described by critical statistics. The latter is an intermediate level statistics  \cite{altshuler1988,shapiro1993} characterized by level repulsion, $P(s)\propto s^2$ for $s \to 0$, as in a disordered metal, but with an exponential, not Gaussian, decay of $P(s)\propto \exp(-\gamma s)$  as in Poisson statistics ($\gamma = 1$) typical of an insulator but 
with $\gamma >1$ \cite{nishigaki1999}. Long-range spectral correlations that describe the time evolution of the system for times shorter than the Heisenberg time have also distinctive features at the transition. The number variance $\Sigma^2$ \cite{mehta2004} is a popular choice to characterize them. It is defined as the variance of the number of levels $N(\epsilon)$ in a spectral interval $\epsilon$. In units of the mean level spacing of the unfolded spectrum
$\left\langle N(\epsilon)\right\rangle = \epsilon \equiv L$ with $L$ the average number of levels in the spectral interval and
\begin{equation}
\Sigma^2(L) = \left\langle N^2(L)\right\rangle - L^2.
\label{eq:nv}
\end{equation}
For quantum chaotic or random matrix ensembles the growth of the number variance is logarithmic, indicating spectral rigidity, for $L \gg 1$. However, for an insulator, it is given by Poisson statistics $\Sigma^2(L) = L$. Around the transition is also asymptotically linear \cite{altshuler1988} but with a slope $\chi < 1$ that depends on the spatial dimensionality \cite{schreiber1996,garcia2007} of the system  ($\chi \approx 0.27$ for the three dimensional case). 

For the sake of completeness, we also compute the spectral form factor, 
 \be\label{eq:sff}
g(t) = \frac{\langle Z^*(t)Z(0)\rangle}{\langle Z^2(0)\rangle}
\ee
with $Z=\sum_ie^{iE_it-\beta E_it}$ with $E_i$ the unfolded eigenvalues and $\beta = 0.001$.
This is an observable which has been employed in recent studies \cite{cotler2016} of the SYK model in the holography literature to detect random matrix like features such as a ramp, a signature of spectral rigidity, for $t$'s of the order of the Heisenberg time.
We do not expect it to posses distinctive features at the transition because the prediction for a similar observable, the power spectrum \cite{garcia2006}, is close to that of an insulator where no ramp is observed because spectral rigidity is absent. Indeed in a disordered insulator $g(t)$ becomes flat after a decay for short times.
\begin{figure}%
	\centering
	\resizebox{0.4\textwidth}{!}{\includegraphics{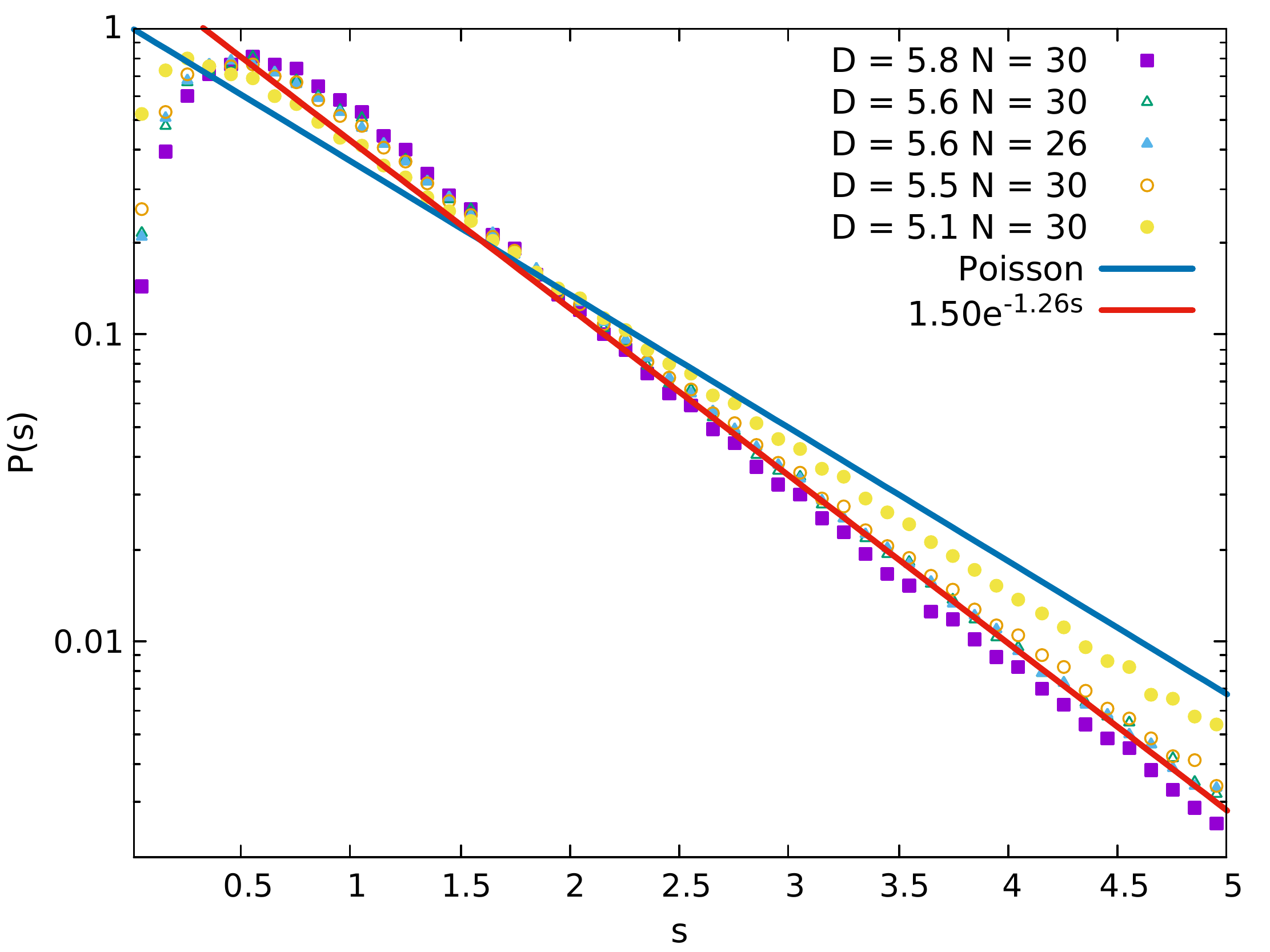}}
	\resizebox{0.4\textwidth}{!}{\includegraphics{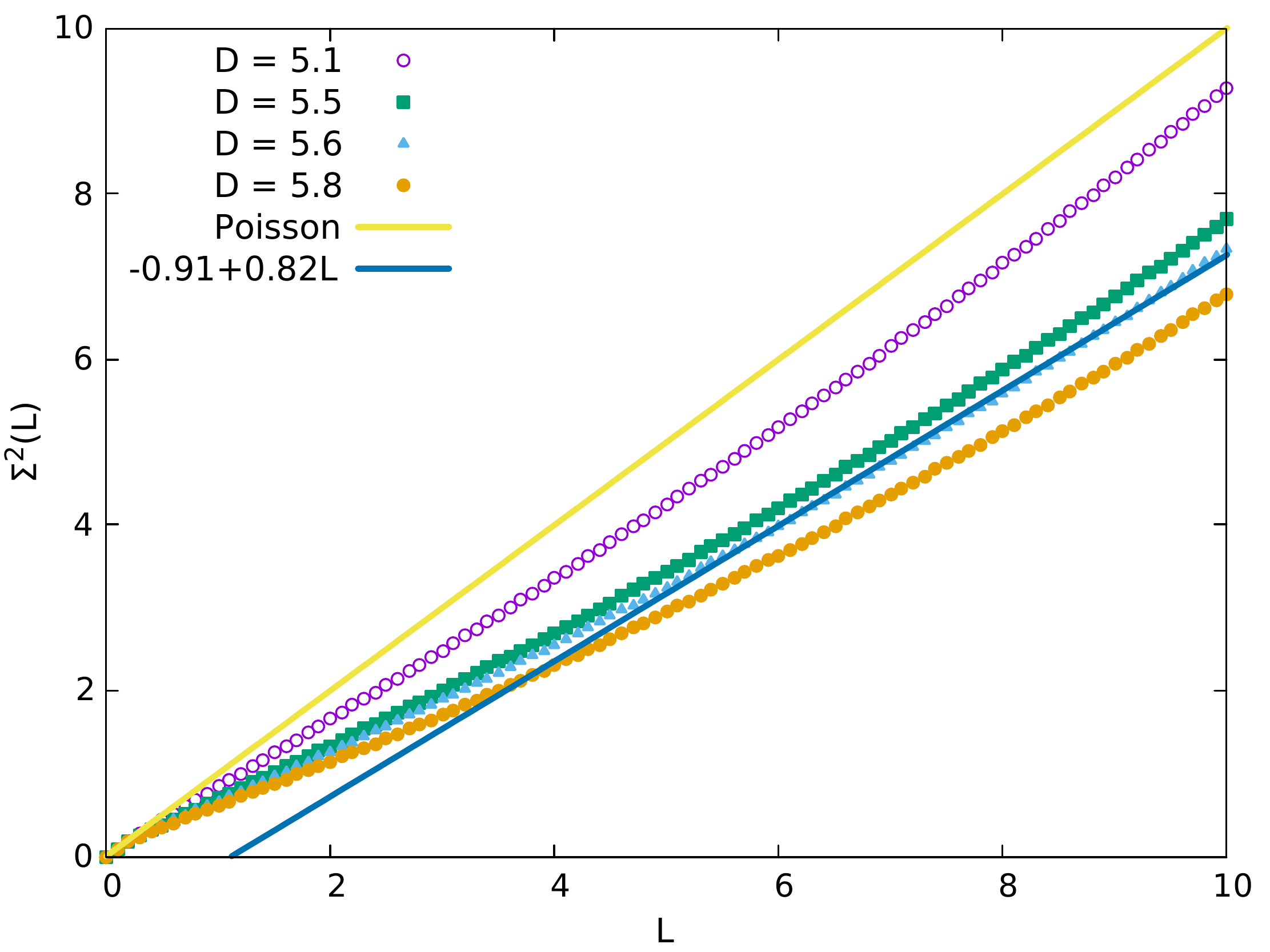}}
		\resizebox{0.4\textwidth}{!}{\includegraphics{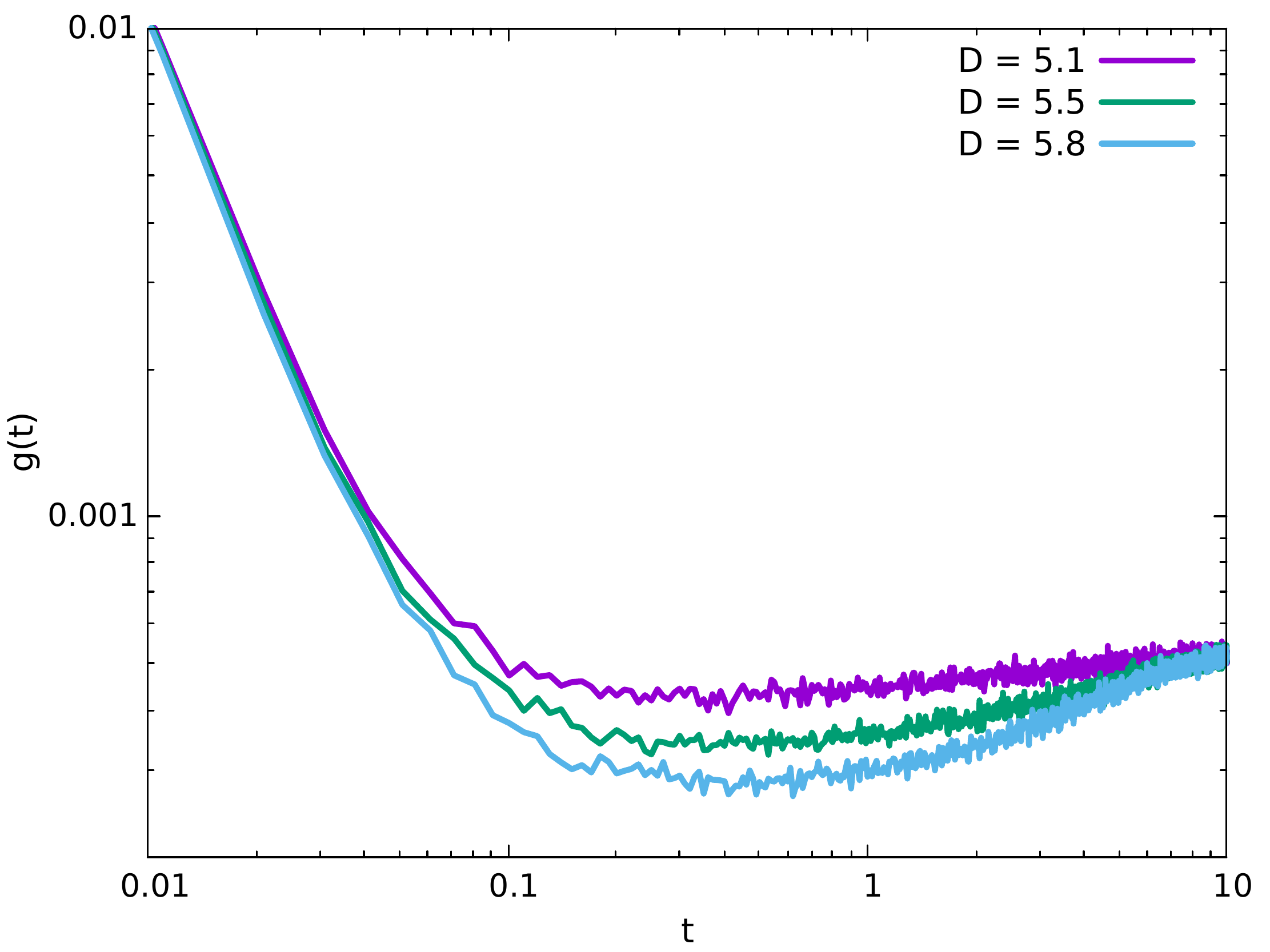}}
	\caption{Upper: $P(s)$ for $d=2$ and different $D \sim D_\mathrm{c} \approx 5.6$. $P(s)$ does not depend on $N$ which is a feature of level correlations at the metal-insulator transition \cite{shapiro1993}. We observe level repulsion, typical of quantum chaos, for $s \ll 1$ while the decay is exponential for $s \gg 1$ as for Poisson statistics though with a different slope. Middle: Number variance $\Sigma^2(L)$ Eq.(\ref{eq:nv}) for $N = 30$. It is asymptotically linear with a slope $\chi \sim 0.8$. Lower: Spectral form factor $g(t)$ Eq.(\ref{eq:sff}) for $N=30$, $\beta = 0.001$ and $D \sim D_\mathrm{c}$. No ramp or dip is observed for $D = 5.1$ which is typical of an insulator while that for $D = 5.8$ both a dip and small ramp start to form. This is consistent with a transition between these two values and with a critical $g(t)$ similar to that for an insulator. The latter is directly related to the linear, instead of logarithmic, growth of the number variance.  All these features have been previously found in systems metal-insulator transitions induced by disorder \cite{shapiro1993,altshuler1988,schreiber1991,bertrand2016}.
	 In all cases, we have taken only the central $20\%$ part of spectrum and the unfolding has been carried out by the splines method that fits locally consecutive subsets of many ($>10$) eigenvalues with low order polynomials.}
	\label{fig:variance1}%
\end{figure}%
In 
Fig.~\ref{fig:variance1} we depict results for the level spacing distribution (top), the number variance (middle) and the spectral form factor (bottom) around the critical point $D = D_\mathrm{c} \approx 5.6$ for different $N$'s. Level statistics share all the features expected of a metal-insulator transition induced by disorder: level repulsion for $s \ll 1$, exponential decay of $P(s)\sim \exp(-\gamma s)$ with $\gamma \approx 1.3 > 1$ and linear number variance with a slope $\chi \approx 0.8 < 1$. As was expected, $g(t)$ is not very sensitive to the transition. It is rather close to that of an insulator with the only difference of the asymptotic value which is hard to determine numerically. However, it is interesting to note that $g(t)$ detects the transition as it has a small dip and ramp for $D = 5.8$, typical of a metal, while for $D = 5.1$ both features are no longer observed, as expected in an insulator.
\section{Outlook and Conclusions} 
Although further investigations are required, our results provide compelling evidence that this generalized SYK model undergoes a metal-insulator transition in the same universality class as the Anderson and many-body localization transition. As a result, it could be relevant as a toy model in studies of the latter and also 
in the exploration of gravity duals that reproduce the phenomenology of many-body localization.  
For instance, it would be interesting to study eigenfunction statistics to confirm whether multifractality of eigenstates \cite{aoki1983,castellani1986,schreiber1991} and a slow approach to thermalization, typical of systems at the metal-insulator transition, also occur in this generalized SYK model.
It would also be worthwhile to explore whether some functional forms of the interaction decay may be amenable to analytical treatment. A necessary condition is likely that off-diagonal replicas \cite{bagrets2016} are negligible. If this is the case, it could still possible to obtain analytical results in the large $N$ limit by solving the associated Schwinger--Dyson mean field equations. 
Preliminary numerical results for power-law and exponential decaying interactions do not show a metal-insulator transition. Therefore it seems that for the transition to occur it is necessary that the interaction vanishes rather abruptly for sufficiently separated Majorana fermions. However it is unclear whether this requirement is an artifact of the small lattice sizes that we can access numerically. 

Another problem that deserves further attention is the study of thermodynamic properties and the Lypunov exponent close to the transition to determine whether are consistent with the existence of a gravity dual. That would be the first step towards the modeling of many-body localization by holographic techniques.

In conclusion, we have found, that the SYK model undergoes a metal-insulator transition in Fock space by reducing the range of the interaction. Level statistics in the (insulating) metallic side are well described by (Poisson) Wigner--Dyson statistics. Around the transition, spectral correlations become $N$ independent and completely agree with those found in  other metal-insulator transition induced by disorder.  

\acknowledgments
MT thanks H. Gharibyan, M. Hanada, S. H. Shenker, and H. Shimada for discussions on spectral analysis during other collaborations.
The work of MT was partially supported by a Grant-in-Aid for Scientific Research on Innovative Areas ``Topological Materials Science'' (KAKENHI Grants No. JP15H05855 and No. JP15K21717), and by Grants-in-Aid No. JP26870284 and No. JP17K17822 from JSPS of Japan.

\bibliography{library2}

\begin{thebibliography}{67}
\bibitem [{not()}]{notefractionalD}%
  \BibitemOpen
  \href@noop {} {}\bibinfo {note} {For integer values of $D$, we have used
  $7680, 864, 200$ samples for $N=22, 26, 30$ respectively, both for $d=2$ and
  $d=\infty$. For non-integer values of $D=5.1, 5.2, \ldots, 5.9$ with $d=2$,
  for each sample, each $\tilde{J}_{ijkl}(D)$ with $l-i=5$ is set to nonzero
  with probability $D-5$, and we have used $768$ samples for $N=22, 26, 30$.
  The error bars have been drawn to indicate the standard error of the mean,
  obtained by dividing the standard deviation of $r$ by the square root of the
  number of the adjacent gap pairs considered}\BibitemShut {NoStop}%
\end{thebibliography}

\end{document}